\documentclass[twocolumn,superscriptaddress,aps,pra,10pt]{revtex4-2}
\usepackage{amsmath}
\usepackage{graphicx}
\usepackage{float}
\usepackage{braket}
\usepackage{amssymb}
\usepackage{xcolor}
\usepackage{siunitx}
\usepackage{tensor}

\usepackage[normalem]{ulem}

\newcommand{\srr}{\hat{n}}

\newcommand{\ddt}{\frac{\mathrm{d}}{\mathrm{d}t}}
\newcommand{\rfac}{r_\mathrm{f}}
\newcommand{\gfac}{\Gamma_\mathrm{f}}

\newcommand{\kay}{\langle k \rangle}
\newcommand{\drfac}{\delta \rfac}
\newcommand{\decay}{\gamma}

\newcommand{\out}[1]{}

\definecolor{patrick}{HTML}{D5B60A}

\begin{document}

\begin{abstract}
Understanding the universal properties of non-equilibrium phase transitions  of spreading processes is a challenging problem. This applies in particular to irregular and dynamically varying networks. We here investigate an experimentally accessible model system for such processes, namely the absorbing-state phase transition (ASPT) of Rydberg-excitation spreading, known as Rydberg facilitation, in a laser-driven gas of mobile atoms. It occurs on an irregular graph, set by the random atom positions in the gas
and, depending on temperature, changes its character from static to dynamic. By studying the behavior of the order parameter in  [\textit{Phys.~Rev.~Lett.}~\textbf{133},~173401~(2024)] we showed numerical evidence for a crossover from directed percolation (DP) universality through various phases of anomalous directed percolation (ADP) to mean-field (MF) behavior when the temperature of the gas is increased. As the behavior of the order parameter is not sufficient to uniquely determine the universality class, we here analyze the distribution of avalanches -- characteristic of non-equilibrium critical behavior --  to fully characterize the ASPT. Performing extended numerical calculations and experiments on a cold $^{87}$Rb atom gas we confirm our earlier numerical findings and our phenomenological model that maps the dynamic network to a static one with power-law tails of the distribution of excitation distances. Furthermore we discuss the influence of dissipation, present in the experiment and a necessary ingredient for the self-organization of the system to the critical point. In particular we study the 
potential modification of the universality class by losses as a function of dissipation strength.
\end{abstract}


\title{Nonequilibrium Universality of Rydberg-Excitation Spreading on a Dynamic Network}


\author{Simon Ohler}
\affiliation{Department of Physics and Research Center OPTIMAS, RPTU Kaiserslautern, 67663 Kaiserslautern, Germany}
\author{Daniel Brady}
\affiliation{Department of Physics and Research Center OPTIMAS, RPTU Kaiserslautern, 67663 Kaiserslautern, Germany}
\author{Patrick Mischke}
\affiliation{Department of Physics and Research Center OPTIMAS, RPTU Kaiserslautern, 67663 Kaiserslautern, Germany}
\affiliation{Max Planck Graduate Center with Johannes Gutenberg University Mainz (MPGC), 55128 Mainz, Germany}
\author{Jana Bender}
\affiliation{Department of Physics and Research Center OPTIMAS, RPTU Kaiserslautern, 67663 Kaiserslautern, Germany}
\author{Herwig Ott}
\affiliation{Department of Physics and Research Center OPTIMAS, RPTU Kaiserslautern, 67663 Kaiserslautern, Germany}
\author{Thomas Niederpr\"um}
\affiliation{Department of Physics and Research Center OPTIMAS, RPTU Kaiserslautern, 67663 Kaiserslautern, Germany}
\author{Winfried Ripken}
\affiliation{Machine Learning Group, Technische Universit\"at Berlin, 10587 Berlin, Germany}
\affiliation{Berlin Institute for the Foundations of Learning and Data, 10587 Berlin, Germany}
\author{Johannes S. Otterbach}
\affiliation{Orthogonal Otter UG, 10961 Berlin, Germany}
\author{Michael Fleischhauer}
\affiliation{Department of Physics and Research Center OPTIMAS, RPTU Kaiserslautern, 67663 Kaiserslautern, Germany}

\date{\today}

\maketitle

\section{Introduction}
The critical behavior at non-equilibrium phase transitions and the phenomenon of Self-Organized Criticality (SOC) \cite{bak1987self, bak1988self, tang1988critical} are closely related to avalanche events - sudden, fast outbursts of energy after longer periods of inactivity.
Although the topic is not without its controversies \cite{watkins201625}, the SOC mechanism is believed to be key to the abundance of real world examples of power-law distributed avalanche-events like earthquakes \cite{sornette1989self}, solar flares \cite{lu1991avalanches, aschwanden201625} and neuron activation in the brain \cite{plenz2021self, hesse2014self}, since it describes how a system can evolve in time to the critical point of a phase transition without an external drive or fine tuning. One of the most important categories of non-equilibrium phase transitions in spreading processes concerns absorbing-state phase transitions (ASPT). Here the behavior of the system changes from an active (spreading) phase with perpetual excitation cascades to an inactive (absorbing) phase, where a single excitation does not change the system at large. Power-law distributed avalanche events are then observed at the critical point between these two phases, reflecting the scale-invariance of the critical state. In this situation, a minimal perturbation can cause a scale-free reaction of the system.

Avalanche events of self-organized critical systems have first been studied in the context of sandpile models, most notably the Bak-Tang-Wiesenfeld (BTW) model \cite{bak1987self, bak1988self}. Shortly after additional systems were proposed that display SOC behavior, such as the Manna- \cite{manna1991two}, Drossel-Schwabl- \cite{drossel1992self} or Olami-Feder-Christensen-model (OFC) \cite{olami1992self} for a different type of sandpile, forest fires and earthquakes, respectively. Most of the avalanche research so far has focused on lattice models, where a toppling or relaxation event is defined as the transfer of e.g. energy or particles to adjacent sites on the lattice. Some works have adapted these models to \emph{networks}, where adjacency is not defined by nearest neighbors on the lattice but instead by links of the graph structure. Here, it was found that the type of graph structure itself can significantly affect the critical behavior \cite{caruso2006olami}.

Additionally, most work so far has been done on \textit{static} systems, where the adjacency relations between individual sites do not change over time. This simplification is justified in many cases, since avalanches typically happen on very short timescales compared to other processes in the system. However, in other systems, for example epidemic spreading, the dynamical properties of the network are very important and happen on relevant timescales \cite{dottori2015sir}. The extent to which graph dynamicity can impact the critical properties of the system is not well understood. In the context of the OFC model, for instance, it was found that the dissipative random-neighbor version results in non-critical behavior compared to critical scaling for fixed connectivity \cite{lise2002nonconservative}, however other authors claim that also in the latter case the model turns non-critical \cite{bonachela2009self}.

Generally, non-equilibrium phase transitions can be associated with a set of critical exponents that describe the scaling of physical observables close to the critical point. For ASPTs, these critical exponents can be related to the exponents of the power-law distributions of avalanche events \cite{hinrichsen2008book}. Obtaining these critical exponents through experiments or simulations is essential to identify the universality class of the ASPT and can help in understanding the relevant physical processes.

For certain ASPTs, however, finding experimental representations can be very difficult. One example is the well-known universality class \textit{directed percolation} (DP), where to this day well-controllable experimental systems are rare, the first one only being discovered in 2007 \cite{takeuchi2007directed}.

Gases of Rydberg atoms offer a versatile experimental platform for the investigation of many-body phenomena, where high-precision measurements on gases as well as on tailored geometries can be performed \cite{browaeys2020many, saffman2010quantum, adams2019rydberg}. Interactions of Rydberg atoms can also be tuned to simulate the dynamics of the SIS (susceptible-infected-susceptible) model \cite{brady2023mean, wintermantel2021epidemic}, which is an important example of a spreading model displaying an absorbing-state phase transition, and Rydberg atoms have been used experimentally to measure avalanche distributions and other critical exponents \cite{helmrich2020signatures}. In this context, the excitation of an atom into a highly excited (Rydberg) state that can spread to other atoms is considered the "active" or infected state, whereas the ground state of the atom is the "passive" or susceptible state.
The spreading of an excitation occurs on a \emph{network} of atoms with fixed spatial separation, given by the so-called facilitation distance. This network can be a regular lattice, if the atoms are trapped e.g. in optical lattice potentials or tweezer arrays, or can be static but random e.g. in a cold gas.

\begin{figure}
    \centering
    \includegraphics[width=\columnwidth]{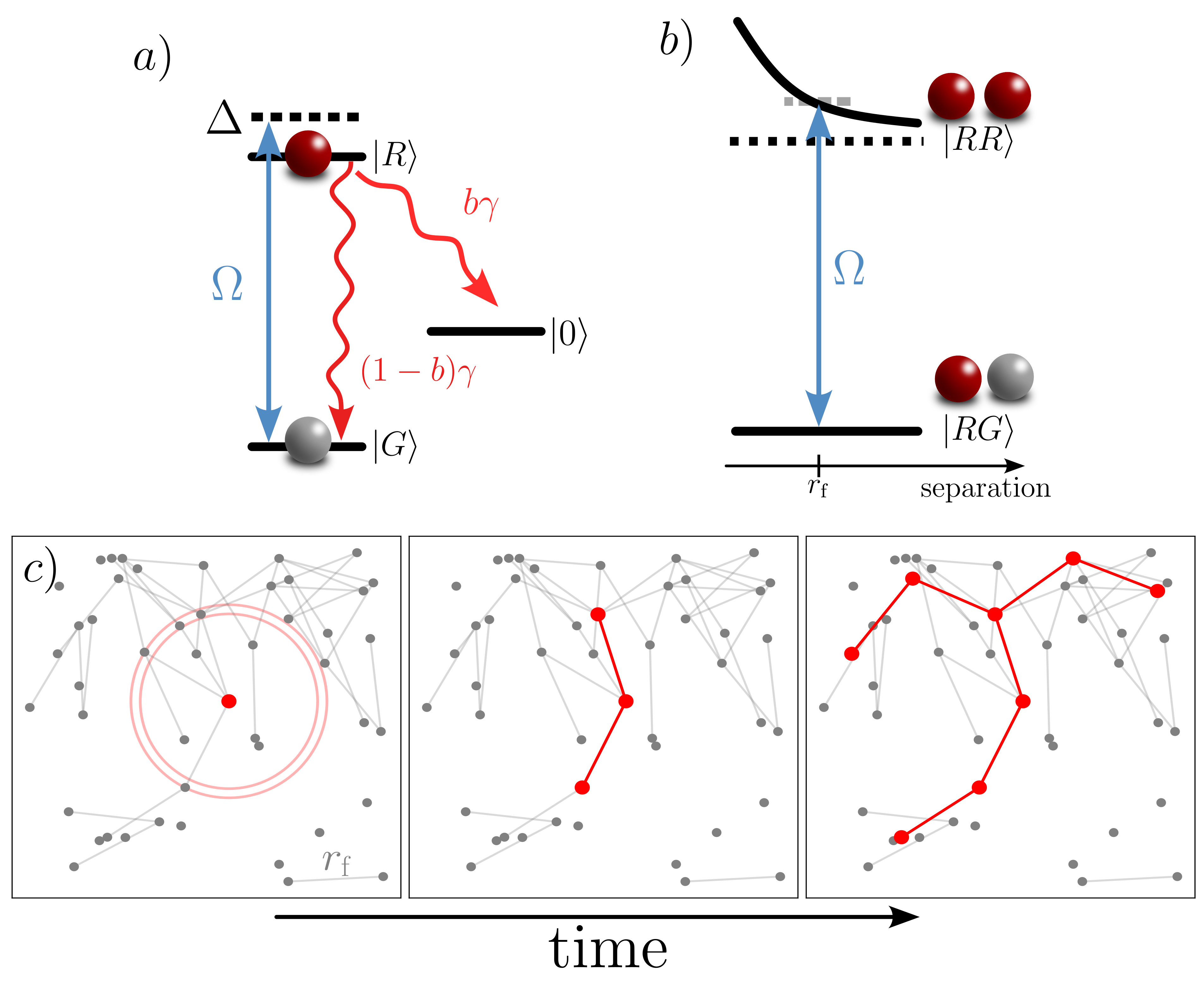}
    \caption{Overview over the microscopic physics in our model. a) Single atom under external drive $\Omega$ with decay channels into ground and inactive states. The parameter $b$ controls the ratio between the decay processes. b) Two atoms with inter-atomic van-der-Waals force that shifts the two-Rydberg state as a function of distance. At $r=\rfac$, the shift cancels the external detuning and the transition becomes resonant. c) Spreading of an avalanche on a network.}
    \label{fig:microscopic_physics}
\end{figure}
An important further aspect of atomic gases is their thermal \emph{motion}. In a recent publication we provided numerical evidence that the ASPT of a driven Rydberg gas under facilitation conditions changes its universality class as a function of the (root mean square) gas velocity \cite{brady2024anomalous}. For low temperatures we obtained DP scaling, changing to anomalous DP (ADP) and eventually mean-field (MF) for higher temperatures. These simulation results explained the unusual experimental measurement value of the critical exponent $\beta$ obtained in a previous publication \cite{helmrich2020signatures}. However, the change in universality was only shown for the critical exponent $\beta$ as well as for one of the correlation length exponents $\nu_{\parallel}$ \cite{brady2024anomalous}, lacking the third critical exponent $\nu_{\perp}$. 
Determining the exponents of the avalanche distribution functions at the critical point provides an alternative way to unambiguously determine the universality class, and we will pursue this approach in the present paper, by both numerical simulations and experimental studies. In addition, the scaling of avalanche critical exponents on dynamical networks such as in Rydberg facilitation remains a mostly open question, only one exponent having been measured in \cite{helmrich2020signatures}.

In this paper, we numerically study the avalanche events in a three-dimensional gas of atoms that are driven by an external laser field and compare the results with experimental data as well as field-theoretical predictions for an effective static model with temperature-dependent power-law tails in the excitation distance. The atomic cloud is characterized by a tunable velocity distribution that, combined with the distance-dependent interaction, yields a dynamical graph on which excitations can spread. As a function of velocity we obtain the avalanche-exponents for area, size and time of the avalanches (for a definition see \ref{subsect:avalanches}) and confirm the universality class crossover from DP to anomalous directed percolation that we found in a previous publication \cite{brady2024anomalous} analyzing the $\beta$ exponent of the order parameter, i.e. the Rydberg density. 
This is a non-trivial result since predicting the avalanche exponents requires knowledge of all three ASPT critical exponents $(\beta, \nu_{\parallel}, \nu_{\perp})$. The numerical results are supported by experimental observations of avalanche distributions of Rydberg facilitation in a cold, trapped gas of $^{87}$Rb atoms. With this, our work also provides the first experimental evidence of ADP universality.

Secondly, the loss mechanisms inherent to self organization of a system to the critical point of an absorbing-state phase transition can affect the universal behavior at criticality or even destroy criticality altogether. For this reason we consider additionally the effect of losses from the excited state and quantify its influence on criticality and exponent values for a frozen gas as well as a finite-temperature gas by numerical simulations. This is especially relevant in the context of our experimental results that invariably include loss.

\section{Rydberg facilitation, model and experimental set-up}

\subsection{Microscopic System}
We study a driven-dissipative system of atoms in three dimensions, where any atom can at any time belong to one of three states, namely the ground state $\ket{G}$, the Rydberg state $\ket{R}$ and the "removed" state $\ket{0}$, which describes a state in which the atom does not take part in the dynamics at all (often called "immune" in the context of epidemic spreading).
We apply an external driving (laser field) with Rabi frequency $\Omega$ that couples $\ket{G}$ to $\ket{R}$ with detuning $\Delta$ (see Fig.~\ref{fig:microscopic_physics}). The Rydberg state can spontaneously decay to the ground state with rate ${(1-b)\decay}$ with $0\leq b \leq 1$. The quantum mechanical evolution of the system can then be described by a Lindblad master equation \cite{lindblad1976generators} for the density operator $\hat{\rho}$, which reads (we set $\hbar=1$)
\begin{align}
    \label{eq:master_equation}
    \ddt \hat{\rho} = i [\hat{\rho}, \hat{\mathcal{H}}] 
    + \sum_l \left(\hat{L}_l \hat{\rho} \hat{L}_l^\dagger
    - \frac{1}{2} \{ \hat{L}_l^\dagger \hat{L}_l, \hat{\rho} \}\right),
\end{align}
where the unitary evolution of the system is given by
\begin{align}\label{eqn:unitary_hamiltonian}
\hat{H} = \sum_i \Omega \hat{\sigma}_i^x - \Delta \srr_i + \sum_{j < i} \frac{c_6}{r_{ij}^6} \srr_i \srr_j.
\end{align}
Here, $\srr_i=(\ket{R}\bra{R})_i$ is the number operator of the Rydberg $\ket{R}$ state, $r_{ij}=|\vec{r_i}-\vec{r_j}|$ is the interatomic distance, $\hat{\sigma}^x$ is the Pauli $x$ matrix. The last term in \eqref{eqn:unitary_hamiltonian} corresponds to the van der Waals interaction between two Rydberg atoms with $c_6$ being the van der Waals coefficient. Using the Lindblad-master equation, the dissipation in the system is taken into account by the Lindblad jump operators ${\hat{L}_1^{(i)} = \sqrt{(1-b)\decay} \, \left(\vert G\rangle\langle R\vert\right)_i
}$, $\hat{L}_2^{(i)} = \sqrt{b\decay} \, \left(\vert 0\rangle\langle R\vert\right)_i$, which describe spontaneous decay of the $i'th$ atom from the Rydberg state into the ground state $\ket{G}$ and the inert state $\ket{0}$, respectively, with the branching parameter $b$. Additionally, we include the effect of dephasing, which stems mainly from laser phase noise and Doppler broadening~\cite{helmrich2020signatures}, but also from the non-zero width of the wave function of the atom over the van der Waals potential \cite{li2013nonadiabatic}, and differential van-der-Waals forces \cite{schlegel2025to}. The dephasing Lindblad-operator reads ${\hat{L}_\perp^{(i)} = \sqrt{\gamma_\perp} \srr_i}$, where $\gamma_\perp$ is the dephasing rate.

In this publication, we always consider the high-dephasing limit of the Rydberg gas, which has been proven to be a good approximation for gaseous Rydberg systems~\cite{ates2006strong}. In this limit, the dynamics of the system is governed by effective rate equations, which can be modeled using a classical Monte-Carlo approach \cite{levi2016quantum} (for more details see Appendix~\ref{sect:appendix_rate_equations}).
%
\subsection{Facilitation Mechanism}
%
The level scheme of a single Rydberg atom and the two-Rydberg dynamics is illustrated in Fig.~\ref{fig:microscopic_physics}. In the \textit{facilitation} regime, the detuning $\Delta$ is chosen sufficiently large to suppress spontaneous (seed) excitations from the ground state. However, if one atom in the system is initially in the Rydberg state, then the van der Waals interaction shifts the Rydberg energy levels of the nearby ground-state atoms. Since the van der Waals interaction is distance-dependent, there exists a distance called the \textit{facilitation radius} $\rfac = (c_6/\Delta)^{1/6}$, at which the van-der-Waals interaction exactly cancels the detuning. In this way, a Rydberg atom can resonantly "pass on" the excitation to other atoms in a spherical shell with radius $\rfac$ around it. The width of this shell $\drfac$ is given by ${\drfac = \frac{\gamma_\perp}{2 \Delta} \rfac}$ with $\drfac / \rfac \ll 1$ \cite{brady2024griffiths}. The rate of the resonant facilitation is denoted by $\gfac=2\Omega^2/\gamma_\perp$, which is an important timescale in the system. Combining these two effects, we see that while an initial (seed) excitation is very unlikely, as soon as Rydberg atoms exist in the system it is possible to observe avalanche-like cascades of excitations. For this a sufficiently high density and strong enough external driving is needed such that the global facilitation rate is stronger than the decay from the Rydberg state to the ground state. At high atom velocities, the Rydberg atoms see a homogeneous ground-state background and the number of facilitated excitations is determined by the density and driving strength only. For a frozen gas with velocity $v=0$, however, the atoms form a network where two atoms are connected if and only if their distance falls into the very narrow interval $\rfac \pm \drfac$. Since the atomic positions are distributed uniformly, the resulting network of atoms that in principle can participate in the Rydberg facilitation is of the Erdős–Rényi \cite{erdHos1960evolution} type \cite{brady2024griffiths}.

\subsection{Rydberg Gas as a Dynamical Graph}

The Erdős–Rényi network for the frozen gas is characterized by a Poissonian distribution of the number $k$ of atoms in the facilitation shell of a single Rydberg atom
\begin{align}
    \label{eq:poissonian}
    P(k) = \frac{(n V_s)^k}{k!} \exp{(-n V_s)}.
\end{align}
Here $V_s\approx 4\pi\drfac\rfac^2$ is the volume of the facilitation shell and $n=n_G + n_R$ the total density of remaining atoms in the ground and Rydberg states. It is well-known that such a network features a percolation transition at the average network degree $\langle k\rangle=1$. To be able to observe universal behavior, the network needs to be above this threshold, since otherwise the system is comprised of disconnected, finite clusters \cite{brady2024griffiths} and a universal data collapse cannot be achieved \cite{brady2024anomalous}. Therefore, the average degree $\braket{k} = n V_s$ needs to be sufficiently larger than unity. To increase $\braket{k}$ in our simulations we increase $n$, however that comes with significantly increased computational cost, leading us to choose $\braket{k}\sim 2.5$ as a reasonable compromise. For this value, approximately $90\%$ of atoms are contained in the largest connected cluster (LCC) of atoms that in principle could undergo Rydberg facilitation.
\begin{figure}
    \centering
    \includegraphics[width=0.9\columnwidth]{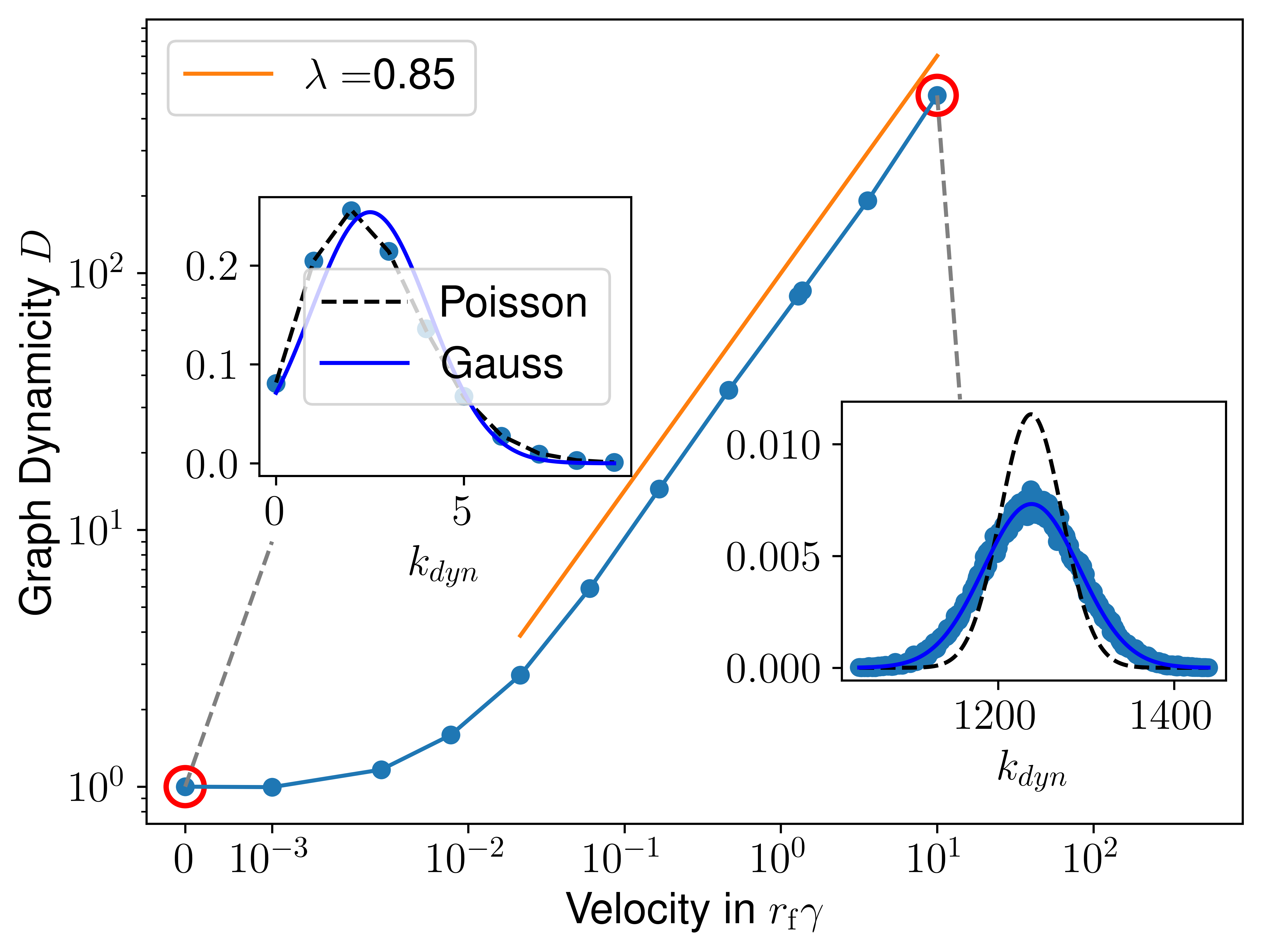}
    \caption{Average number of unique links an atom encounters over the time $1/\decay$ as a function of rms velocity normalized to the $v=0$ case. The insets show the full distribution for velocities $v=0$ and $v=10\rfac\decay$ as well as Poissonian and Gaussian fits.}
    \label{fig:k_ttl_over_k_inst}
\end{figure}
For finite temperature, the system has to be represented as a dynamical graph, since the distances between the atoms change over time. This implies that the pairs of atoms between which an excitation can spread (pairs of atoms with a mutual distance close to $\rfac$) change over time, which, in the language of graph theory, corresponds to the creation and desctruction of links between nodes. Changing the gas velocity then allows to change the rate of link creation and destruction and therefore the degree to which dynamical effects become relevant. We quantify the degree to which the graph is dynamical by introducing the graph dynamicity $D$ which reads
\begin{align}\label{eqn:graph_dynamicity}
D = \frac{\braket{k}_{dyn}}{\braket{k}_{stat}},
\end{align}
where $k_{dyn}$ counts all unique atoms (nodes) that an atom has ever been connected to during the inverse decay time $1/\decay$ and $k_{stat}$ is the average number of instantaneous connections (determined fully by the density and width of the facilitation shell). The result can be seen in Fig.~\ref{fig:k_ttl_over_k_inst}. We observe that after a period of slow growth, starting from $v\sim0.02\rfac\decay$ we see a continuous power-law increase in the number of unique connections. Additionally, at this point the distribution of unique partners changes from Poissonian (low-velocity) to Gaussian (high-velocity). Coincidentally, this velocity scale agrees well with the upper limit of DP universality found in \cite{brady2024anomalous}. In addition to the data we also show the power-law fit to the velocity interval from $v>0.06\rfac\decay$, which yields the exponent $\lambda=0.85$. The relevance of this power-law increase in dynamicity is however unclear. We note that this consideration takes into account solely the dynamical graph structure of atoms on which Rydberg facilitation is in principle possible, not the actual excitation dynamics that depends on other factors like the external drive intensity.

\subsection{Experimental setup}

\begin{figure}
    \includegraphics[width=\columnwidth]{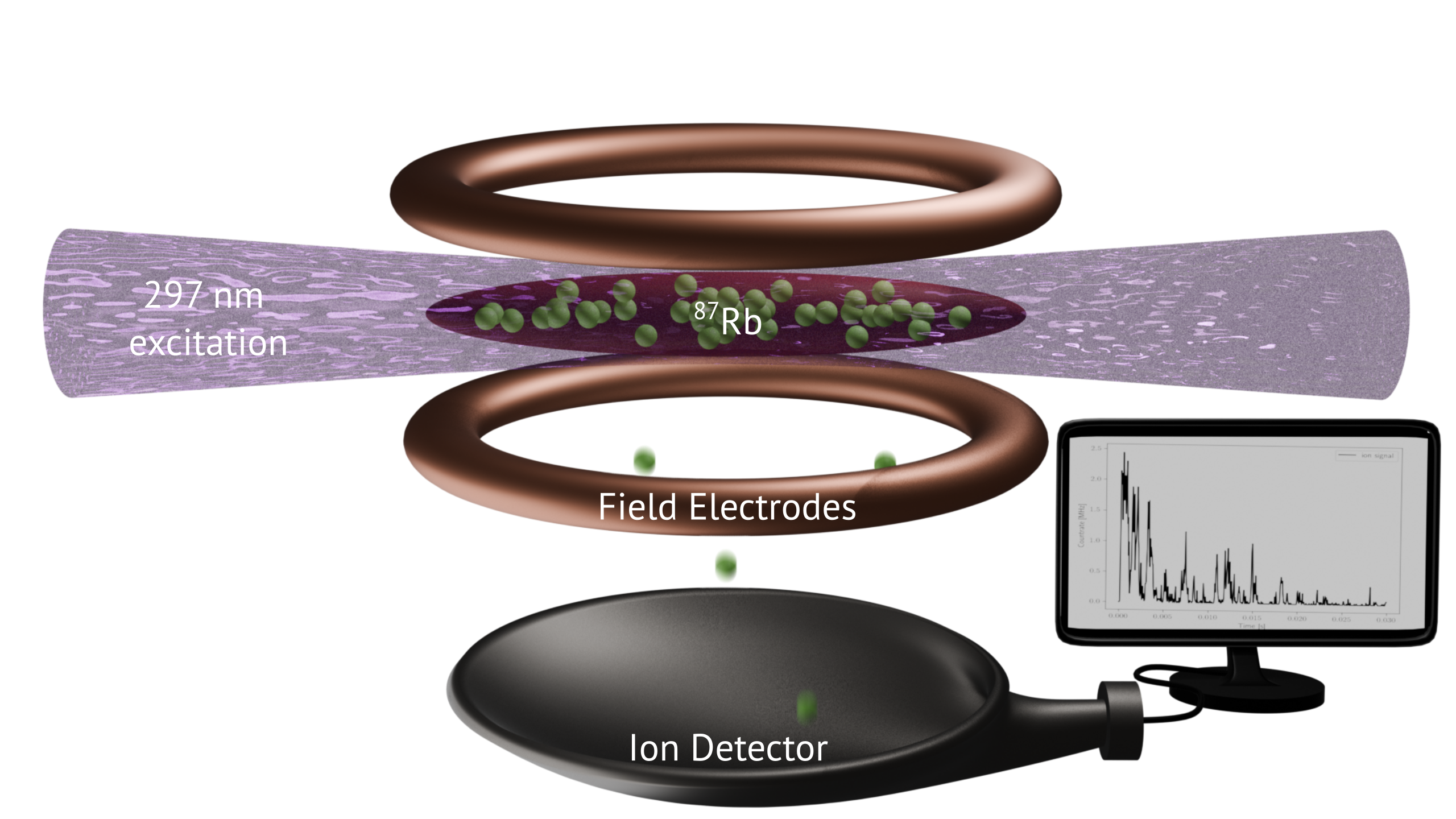}
    \caption{Sketch of the experimental setup. The Rydberg laser off-resonantly drives facilitation dynamics and cascades of Rydberg excitations (green circles) appear in the cold cloud. Rydberg atoms can decay into ions via photo- or associative-ionization, which are guided to an ion detector where their arrival time is recorded. This allows to observe the facilitation dynamics continuous in time.}
    \label{fig:experiment_sketch}
\end{figure}

To experimentally study the collective Rydberg facilitation dynamics, we prepare a cloud of $^{87}\text{Rb}$ atoms in a crossed optical dipole trap with trapping frequencies of $\omega_{x,y,z} = 2\pi \times (332, 332, 73)\,\mathrm{Hz}$. The experimental setup is sketched in Fig.\ref{fig:experiment_sketch}. Forced evaporation in the dipole trap is stopped at a final temperature of $T=\SI{1}{\micro\kelvin}$, leading to a thermal cloud with a  density of $\rho = 2.2\cdot10^{13}\,\mathrm{cm}^{-3}$. This temperature corresponds to  an rms velocity of ${v=0.39\pm0.26\rfac\decay}$. The facilitation dynamics is induced by off-resonant, blue-detuned ($\Delta = \SI{40}{\mega\hertz}$), continuous excitation from the $\ket{G}\equiv\ket{5S_{1/2}~F=2~m_F=2}$ ground state to the $\ket{R}\equiv\ket{40P_{3/2}}$ Rydberg state for a total duration of $\SI{100}{\milli\second}$.
Due to photoionization from the dipole trap lasers and associative ionization, a fraction of the Rydberg population gets ionized.
This decay channel contributes to the dissipation process that brings atoms into the inactive state $\ket{0}$. This state accounts for ions, atoms lost from the system since the Rydberg state is not trapped, and also atoms that decayed to the $\ket{5S_1/2~F=1}$ ground state which does not participate in the excitation dynamics.  We estimate that approximately $2/3$ of the Rydberg excitations decay back to the $\ket{G}$-state, i.e. the branching ratio is $b\approx 0.3\pm 0.15$. The created ions are accelerated through a small electric field towards an ion detector where their arrival time is detected.
In this way, we obtain a time-continuous measurement signal proportional to the Rydberg density, which allows the observation of the facilitation dynamics in situ. Due to the discrete nature of the ion arrival information, binning the data is required to obtain an ion rate.
%

\section{Universality class and Avalanche Distribution}

\subsection{Crossover of Universality Classes in Rydberg Facilitation}
%
\begin{figure}
    \includegraphics[width=\columnwidth]{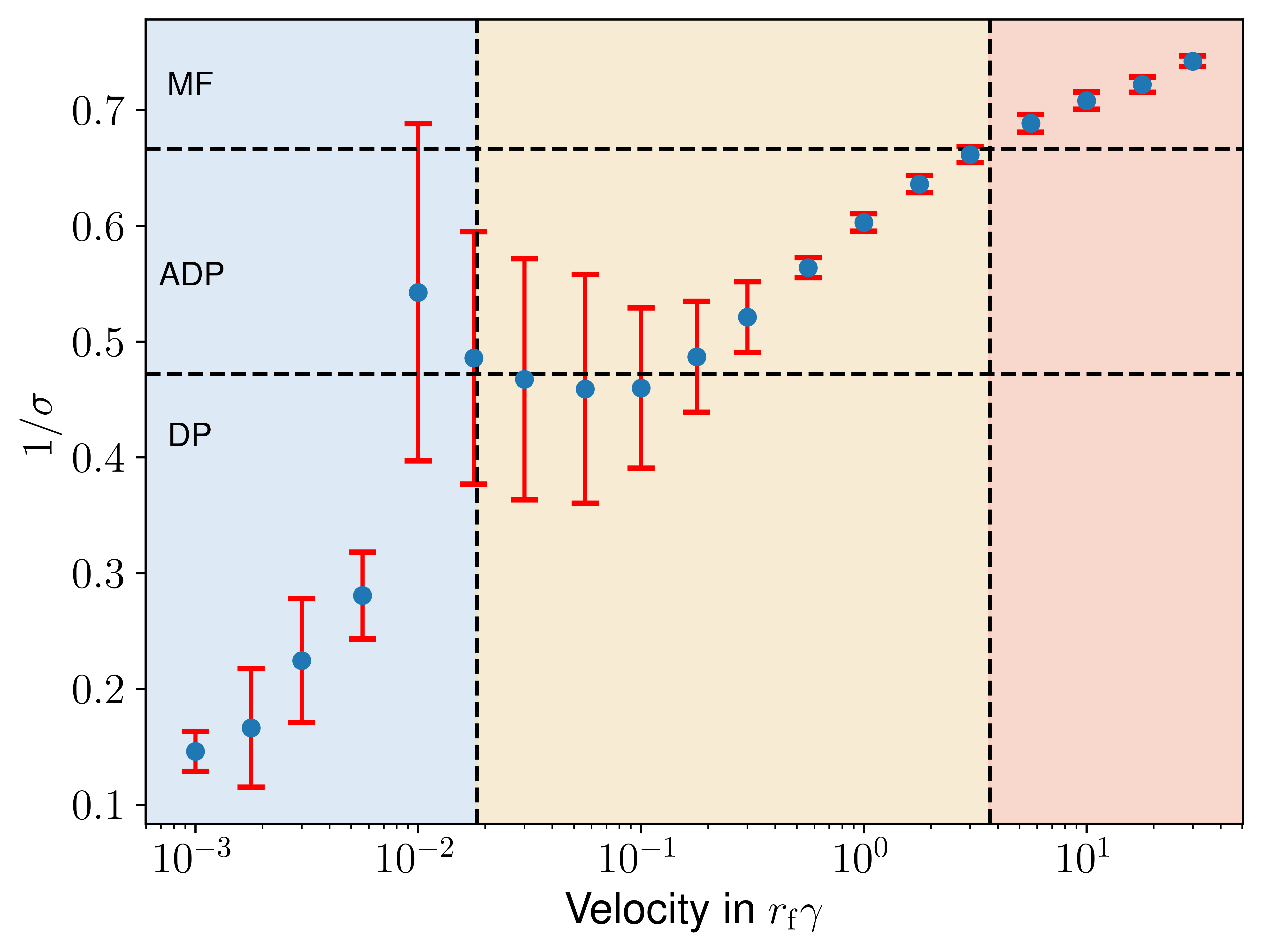}
    \caption{Reciprocal Lévy-flight parameter as a function of gas velocity. The horizontal lines denoting the transition from one universality class to another are given by \cite{hinrichsen2008book}. The vertical lines correspond to the velocities $v_{-}=\drfac\gfac$ and $v_{+}=\rfac\gfac$. Shown data has also been presented in \cite{brady2024anomalous}.}
    \label{fig:sigma_over_v}
\end{figure}
In \cite{brady2024anomalous}, it was shown by extensive numerical simulations that the absorbing-state phase transition in a Rydberg gas changes its universality class from DP through ADP to MF by varying the velocity of the atoms. This crossover was explained by mapping the Rydberg facilitation dynamics on the dynamic network to a spreading process on a fixed network, however with long-distance power-law tails in the distribution of excitation distances. The power-law tails emerge when considering the probability that a moving Rydberg atom excites the next atom at a minimum distance $r= J\delta z$ but ignoring the associated time delay in excitation, where $\delta z$ is some small length interval and $J$ is an integer \cite{brady2024anomalous}.
This probability reads ${P(X > J) = (1 - p_\mathrm{exc})^J}$, where $p_\mathrm{exc}$ is the excitation probability in a given infinitesimal time interval ${\delta t = \delta z / v}$. For a single atom the excitation probability in this interval $\delta t$ is given by ${p_\uparrow = 1 - \mathrm{e}^{-\gfac \delta t}}$, 
and since the number of atoms in the facilitation shell of a Rydberg atom is  Poissonian distributed, the excitation probability reads
$p_\mathrm{exc}= \sum_{k=0}^{\infty} P(k) (1 - (1 - p_\uparrow)^k)$, i.e.
\begin{equation}
        p_\mathrm{exc} 
        \label{eq:p_exc}
        =  1 - \exp\Bigl\{-\xi \delta z\Bigr\},
\end{equation}
with ${\xi = \frac{\kay}{\drfac} (1 - \mathrm{e}^{-\drfac \gfac / v})}$. This then yields the probability distribution of distances  $r$  from the initial position of the Rydberg atom for the first successive excitation
 \begin{align}
        P(r) = 
        \label{eq:f_of_r}
        &2 \pi \xi r \int_0^\pi \!\!\!\mathrm{d} \theta \, \, \frac{\mathrm{e}^{-\xi(\sqrt{\cos^2\theta + r^2 - 1} - \cos\theta})}{\sqrt{\cos^2\theta + r^2 - 1}},
    \end{align}
%
where ${\vec{r}= \rfac \hat{e}_r(\theta, \varphi) + (0, 0, z)^T}$. 
We have shown in \cite{brady2024anomalous} that this distribution agrees very well with the numerically obtained distribution of first excitations in the finite-temperature gas. It also agrees very well with a power-law fit of the form
\begin{align}
    P_{\mathrm{hop}}(r)\sim \frac{1}{r^{d+\sigma}},
\end{align}
which resembles a Lévy-flight statistic for $d$ being the dimension of space and $\sigma$ the Lévy-flight parameter.

In Fig.~\ref{fig:sigma_over_v} we show the inverse Lévy-flight parameter over the gas velocity $1/\sigma$, obtained from power-law fits to the distances between excitations in the simulated gas \cite{brady2024anomalous}. We see that the Lévy-flight parameter $\sigma=\sigma(\overline{v})$ depends on the average velocity $\overline{v}= (\langle v^2\rangle)^{1/2}$ of the atoms, causing a transition between universality classes. The critical exponents $(\beta,\nu_\parallel,\nu_\perp)$ characterizing the behavior of the order parameter (Rydberg density), temporal and spatial correlations, respectively, can be approximated as a function of $\sigma$ close to the MF regime via a renormalization-group approach, where for $d=2\sigma - \epsilon$, $\epsilon$ being a small parameter, the exponents can be written as \cite{hinrichsen1999model}
\begin{align}
    \beta & = 1 - \frac{2\epsilon}{7\sigma} + O(\epsilon^2),\nonumber
    \\
    \nu_\perp & = \frac{1}{\sigma} + \frac{2\epsilon}{7\sigma^2} + O(\epsilon^2),  \label{eq:RG}
    \\
    \nu_\parallel & = 1 + \frac{\epsilon}{7\sigma} + O(\epsilon^2).\nonumber
\end{align} 
In \cite{brady2024anomalous} we determined two of the three critical exponents $\beta$ and $\nu_\parallel$, characterizing the order parameter and temporal correlations, while the third one, $\nu_\perp$, which determines spatial correlations, was not accessible. In the following we discuss an alternative approach to fully determine the universality class, which we pursue in this work, both numerically and experimentally.
\ 

%
\subsection{Avalanches}\label{subsect:avalanches}
%
A characteristic phenomenon in dynamical systems close to the critical point of an ASPT is the appearance of avalanches, i.e. cascades of excitation events spreading through the system.
Their time $t$ (duration), area $a$ and size $s$ are random but show a power-law probability distribution, which reflects the scale invariance at the critical point. For the definitions of area and size we follow \cite{hinrichsen2008book}, where the area is the number of unique sites (in our case atoms) that were involved in the avalanche, while size is the total number of relaxation (in our case decay) events that took place in the avalanche, counting possibly multiple relaxations for a single atom. Time is measured from the first excitation to the last decay of the avalanche. The distributions then scale as
\begin{align}
    P(t) &\sim t^{-\tau_t},\nonumber\\
    P(s) &\sim s^{-\tau_s},\\
    P(a) &\sim a^{-\tau_a}.\nonumber
\end{align}
Generally, the critical exponents of the ASPT $(\beta,\nu_\parallel,\nu_\perp)$ are connected to these avalanche exponents in the following way \cite{hinrichsen2008book}
\begin{align}\label{eqn:tau_of_beta}
\tau_a = 1 + \frac{\beta}{d\nu_{\perp}},
&&
\tau_s = 1 + \frac{\beta}{\nu_{\parallel} + d\nu_{\perp}-\beta},
&&
\tau_t = 1 + \frac{\beta}{\nu_{\parallel}}.
\end{align}
For the MF case, the critical exponents $\beta,\nu_\perp,\nu_\parallel$ are known exactly, while for DP there exist numerical estimates in the literature \cite{munoz1999avalanche}. For the case of ADP, the critical scaling of the system depends on the 
Lévy-flight parameter $\sigma=\sigma(\overline{v})$.

\subsubsection{Simulation Results}
To numerically investigate the critical properties at the boundary between absorbing and active phase in the Rydberg gas, we first need to pin-point the parameters to obtain the critical state, namely the critical driving-strength parameter $\Omega_c$ and the corresponding critical density $n_c$. To see universal scaling in our system, the underlying Erdős–Rényi-network needs to be well above the percolation threshold \cite{brady2024griffiths}, so we fix $n_c=20.0\rfac^{-3}$, which corresponds to an average network degree of $\langle k \rangle=2.5$. This holds for all gas velocities that we consider, while $\Omega_c$ changes as a function of velocity. We then determine the critical driving strength $\Omega_c$ via active-density decay, where we start with all atoms in the excited state and observe the decay process to the ground state $(b=0)$ (see also Appendix \ref{appendix_critical_point_estimation}). In the absorbing state, the active density $\rho$ will decrease exponentially with time, for an active state $\rho$ approaches a constant value and for the critical state we expect power-law decay \cite{munoz2010griffiths, brady2024griffiths}. Note that we could simply use the SOC mechanism ($b\ne 0$) to obtain the critical state by starting in the active phase and time-evolving to the phase transition. However, to achieve sufficient accuracy this would necessitate an extreme separation of time scales which is computationally more expensive.

\begin{figure}
    \centering
    \includegraphics[width=\columnwidth]{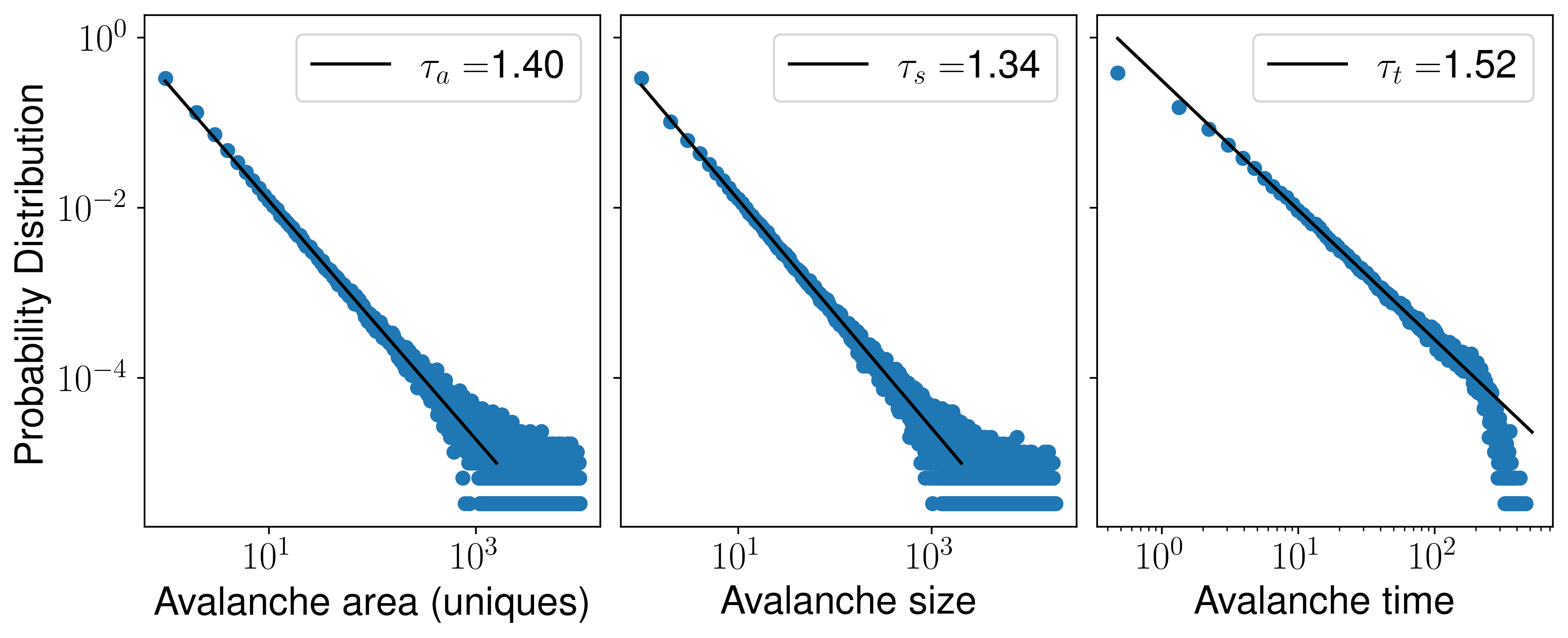}
    \caption{Simulation avalanche area, size and time distributions for $v=0$ and a system size of $L=20\rfac$. The power-law fit function is shown in black.}
    \label{fig:distributions_v_0}
\end{figure}

Having obtained the critical parameters, we perform repeated calculations where we generate a gas of ground state atoms at density $n=n_c$ and driving $\Omega=\Omega_c$. Furthermore, we use $\Delta/\decay=1000$, $\gamma_\perp/\decay=20$ and a cubic simulation volume with edge length $L$ and periodic boundary conditions. During a simulation, we place a single excitation in the gas and let the system evolve until no excited atoms remain, extracting the area, size and time of the avalanche. In this manner, we obtain approximately $10^5$ avalanches for each configuration of parameters, allowing us to analyze the occurrence statistics. See Fig.~\ref{fig:distributions_v_0} for example distributions for a fixed system size at $v=0$. We do not extract the avalanche exponents from such distributions directly, but perform a finite-size expansion on the fitted exponents for system sizes up to $L=20\rfac$. For more details see Appendix \ref{appendix_fse}.

Using the two sets of equations, eq.\eqref{eq:RG},\eqref{eqn:tau_of_beta} and the mapping $\sigma(v)$ from \cite{brady2024anomalous} (see Fig.~\ref{fig:sigma_over_v}) we can make predictions about the expected avalanche exponents $\tau_t, \tau_a, \tau_s$ over the Rydberg gas' velocity in the DP, ADP and MF regimes. In \cite{brady2024anomalous} we additionally found that the Rydberg gas enters the ADP II phase in between DP and ADP, which is additionally characterized by Lévy-flight distributed waiting times between facilitation (infection) events. For this regime, however, we cannot obtain theoretical predictions for the critical exponents.

\begin{figure*}
    \centering
    \includegraphics[width=\textwidth]{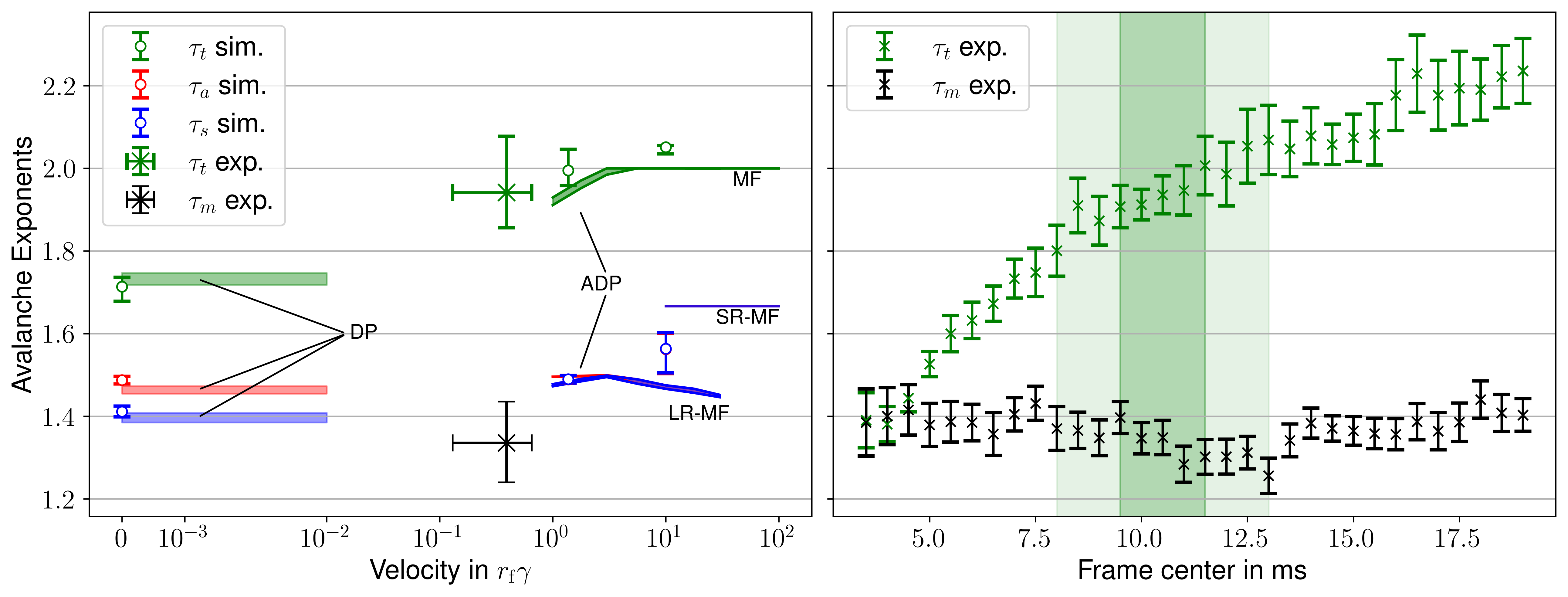}
    \caption{Left plot: Simulation avalanche exponents over velocity (in units of $\rfac\decay$) for $b=0$. Each velocity uses a different value of $\Omega=\Omega_c$. Note that for both mean-field cases, the theoretical predictions for area and size become identical. Also shown are the experimentally extracted values for time and magnitude (in the experiment area and size cannot be distinguished). For an explanation of the latter, see text. Experimental values are derived from the highlighted area in the right plot. Right plot: Experimental avalanche exponents as a function of time. Error bars represent the fit uncertainty, additional systematic errors may arise from e.g. binning. The dark green area corresponds to our estimate for $t_\mathrm{crit}$, at which the critical point of the ASPT is reached. For more details on the $t_\mathrm{crit}$ estimation see Appendix~\ref{appendix_critical_point_estimation}}.
    \label{fig:tau_over_velocity}
\end{figure*}

In Fig.~\ref{fig:tau_over_velocity} we show the predictions in the DP, ADP and MF phases combined with both the results of the simulations as well as the experimental data for the time and magnitude exponents. The experimental magnitude exponent corresponds to the area and size exponents, see the next section for details. For the DP values, the thickness of the bars indicate the uncertainty of the avalanche exponents derived from the uncertainty of the critical exponents  $(\beta, \nu_\perp, \nu_\parallel)$ in the literature. For the ADP values, the uncertainty interval is computed from the $\sigma$ uncertainty as shown in Fig.~\ref{fig:sigma_over_v}.
We observe that for the expected DP and ADP universality classes the simulation results agree very well with the literature values and the ADP values found in \cite{brady2024anomalous}. For the case of $v=10\rfac\decay$ in the MF regime we see that while the time exponent remains close to theoretical predictions, the area and size exponents are larger than the expected long-range (LR) MF case. 

There are two distinct MF cases which result in different area and size exponents, depending on the dimension and the range of interactions \cite{hinrichsen2008book}. Short-range (SR) MF is expected for $d\geq 4$ and $\sigma\geq 2$ and is characterized by the critical exponent $\nu_\perp^{\textrm{SR}}=1/2$. In contrast, the LR MF case appears for arbitrary dimension as long as $\sigma<\min(2, d/2)$ and is associated with the value $\nu_\perp^{\textrm{LR}}=1/\sigma$. Our simulations give values of the area/size exponents which are in between the SR and LR mean-field predictions. We do not have an understanding for this behavior. However, it is important to note that both the SR-MF as well as the LR-MF literature values were obtained from static models on regular lattices, whereas we consider a dynamical graph. Obviously the mapping of moving atoms with close-range interactions to a static network with long-range connections (Lévy-Flights) breaks down when the atom velocity becomes too large.

\begin{figure*}[t]
    \includegraphics[width=\textwidth]{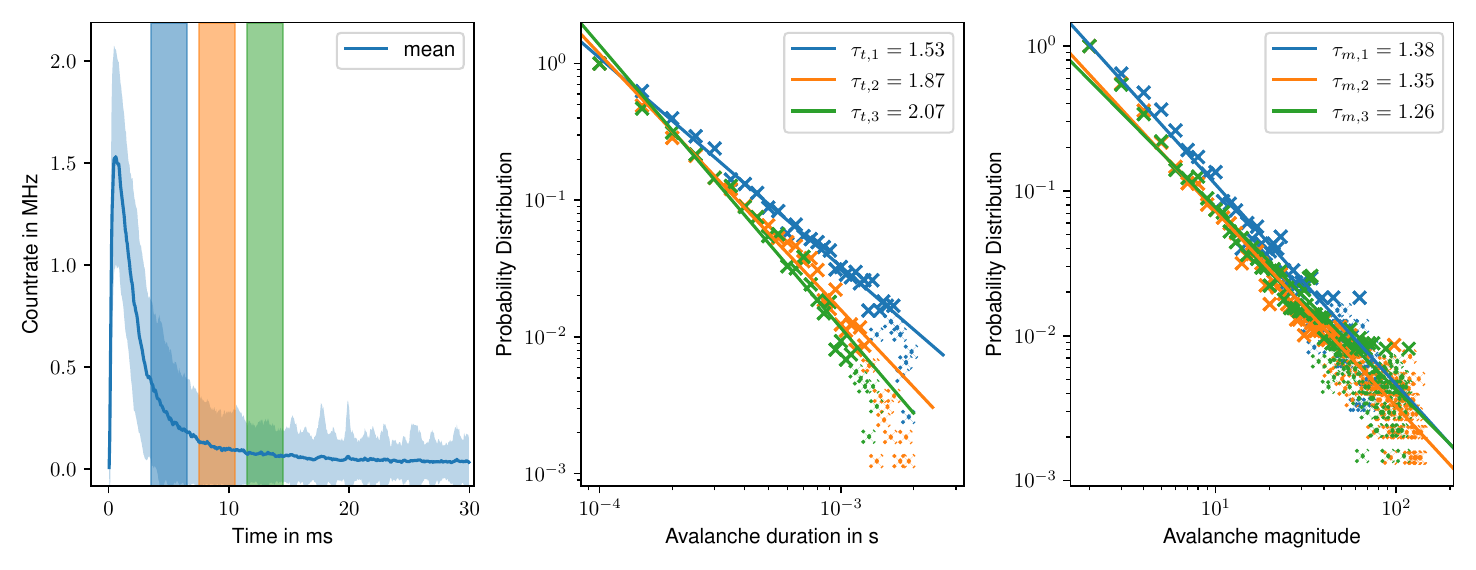}
    \caption{Left: Average ion countrate during the excitation pulse, averaged over 1000 realizations (blue line). The dynamics starts in the active phase. The shaded blue area illustrates the standard deviation between runs. The colored vertical boxes show the $\Delta t_\mathrm{window}=\SI{3}{\milli\second}$ long time windows used to extract the avalanche exponents (centered at blue: 5 ms, orange: 9 ms, green: 13 ms). Middle/Right: Normalized distribution of avalanche durations/avalanche magnitudes in the three different time windows (blue, orange and green). Dashed data points have been ignored for fitting the algebraic distribution.}
    \label{fig:experiment}
\end{figure*}

\subsubsection{Experimental Results}
In order to observe the system at the critical point in the experiment, the gas is initialized with parameters for the density and driving strength such that the dynamics always starts in the active phase. Due to atom loss, as described above, the density and thus the effective driving strength are reduced and the system evolves towards the critical point (self-organisation of criticality). In the vicinity of the critical point the detected ion signal distribution becomes clearly non-Poissonian and avalanches of various sizes can be observed. The atom loss rate is reduced as it scales with the Rydberg density. 
After reaching the critical point, the system evolves slowly away from the critical point for two reasons. Firstly the off-resonant driving laser continues to create excitations at a low rate even in the absorbing phase. Secondly as long as there are Rydberg excitations present, atoms will continue to become ionized or decay into inert states. In the absorbing phase, the system, even though it is non-critical, still shows algebraic behavior over finite scales. This dynamic in the measurement poses two challenges when extracting the critical avalanche exponents: i) We need to find a reliable method to extract avalanches in a system where the starting and ending time of avalanches are masked by a random seed process and ii) we have to estimate rather accurately where the critical point is reached during the time evolution of the sample. 
To distinguish different avalanches, we bin our data in $t_\mathrm{bin}=\SI{50}{\micro\second}$ intervals corresponding to the theoretical lifetime of the $40P_{3/2}$ Rydberg state and consider an avalanche to end and the next avalanche to start if one of those bins does not contain any counts.
The duration of an avalanche is then given by the number of consecutive nonempty bins, and the magnitude of the avalanche is quantified by the number of events in those bins.
It is not possible to count the number of times an atom gets excited to the Rydberg state and decays back to the ground state, and at the same time, not all atoms that have been excited end up ionized.
Therefore, the experimental magnitude of an avalanche cannot be exactly mapped to either the area or size exponent.
However, the difference in theoretical prediction between both exponents in the ADP regime is negligible compared to the experimental uncertainty.

When the system evolves from the active to the absorbing state, it eventually becomes critical.
It is, however, not trivial to precisely determine at what time $t_\mathrm{crit}$ the critical point is reached as can be seen by looking at the average ion count rate during the measurement shown in the left plot of Fig.~\ref{fig:experiment}. Starting in the active phase, we observe a steep drop in the signal \SI{2}{\milli\second} after the pulse started, corresponding to a rapid reduction in the number of facilitation partners due to loss into the inert state.

In contrast, after \SI{15}{\milli\second} the strongly reduced decay rate signals the absorbing phase where the decay is driven by the off-resonantly created seeds and the exponentially dying avalanches.
The distribution of avalanche durations and magnitudes remains algebraic around the critical point as shown in Fig.~\ref{fig:experiment}.
To account for the uncertainty in the precise time $t_\mathrm{crit}$ at which the critical point is reached, we analyze the avalanche exponents in three overlapping time windows of length $\Delta t_\mathrm{window} = \SI{3}{\milli\second}$ in the time range \SI{3.5}{\milli\second} to \SI{14.5}{\milli\second}.
By fitting an algebraic function to the avalanche occurrence statistics in the respective time window we can extract the time and magnitude exponents for the avalanches.
Since the algebraic behavior prevails even away from the critical point, the presence of algebraic behavior alone cannot serve as indicator for the time at which the system is critical. 
Instead, to estimate $t_\mathrm{crit}$, we analyze the activity distribution in the corresponding time windows as detailed in Appendix \ref{appendix_critical_point_estimation} and obtain an estimate of $t_\mathrm{crit}\approx\SI{8}\dots\SI{13}{\milli\second}$ (light green in Fig.~\ref{fig:tau_over_velocity}), i.e. centers of the overlapping evaluation windows ranging from \SI{9.5}{\milli\second} to \SI{11.5}{\milli\second} (dark green in Fig.~\ref{fig:tau_over_velocity}).

In Fig.~\ref{fig:tau_over_velocity} we plot the results for the simulation along with the data extracted from the experiment. The horizontal error bars in the left plot denote the uncertainty regarding the lifetime of the Rydberg states. In the right plot, the continuously changing avalanche exponents are shown. 
Additional contributions to the uncertainty in the exponent values emerge from systematic sources such as the choice of $t_\mathrm{bin}$ and are not shown.

We can see that the experimental time exponent for the critical interval falls into the ADP range and is clearly incompatible with the DP universality class. The magnitude exponent is smaller than expected for ADP universality, however it is unclear if this deviation might be caused by our measurement imperfections.

Additionally, the experimental data allows to exclude other, similar universality classes like the Manna class \cite{manna1991two}, where in 3D a time exponent of $\tau_t^{Manna}\approx 1.78$ is expected. We can also rule out that the Rydberg gas belongs in the BTW-class universality, as for that model a 3D time-exponent of $\tau_t\sim 0.92$ was predicted \cite{bak1988self}. However, one should note that the 2D predictions of said Ref.~\cite{bak1988self} are subject of intense debate since no simple finite-size scaling seems to exist \cite{manna1990large, lubeck1997numerical, hinrichsen2008book}.

\section{Influence of Dissipation}\label{sect:dissip_avals}

\subsection{Effect of self-organization on Erdős–Rényi character of excitation graph}

\begin{figure}[H]
    \centering
    \includegraphics[width=0.9\columnwidth]{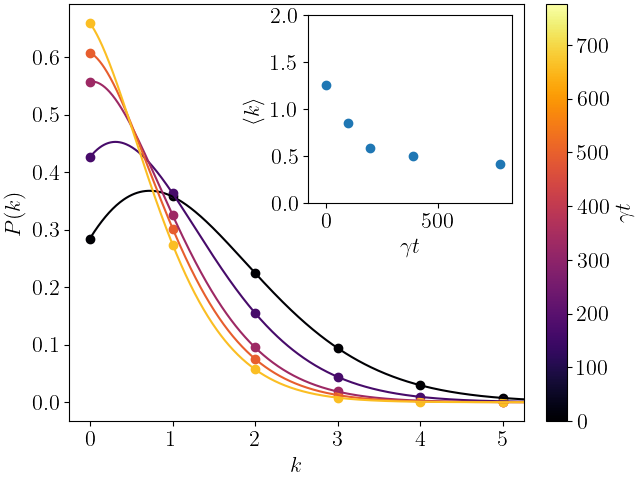}
    \caption{Node degree distribution (dots) for different times in the frozen gas with loss parameter ${b = 0.3}$, and Poissonian fit with fit parameter $\braket{k}$ (solid lines). Fit parameter (average degree) $\braket{k}$ over time (inset).}
    \label{fig:p_of_k}
\end{figure}

One important question is how the decay channel $b\decay$ into the inert state $\ket{0}$, responsible for the self-organization of the facilitated Rydberg gas to the critical point,  affects the properties of the network of possible excitations. As atoms in the inert state no longer interact with other atoms, we do not consider them as part of the graph. Furthermore since the decay into the inert state affects only atoms in the Rydberg state it may lead preferentially to a loss of large clusters of atoms which are pairwise in facilitation distance. This could affect the structure of the network in particular  for a frozen gas, where this network is a static Erdős–Rényi network. To analyze this effect, we calculate the degree distribution $P(k)$ for an initially percolating graph with loss parameter ${b=0.3>0}$ (see Fig.~\ref{fig:p_of_k}).

By fitting with a Poissonian function, with fit parameter $\braket{k}$, we find that the degree distribution $P(k)$ remains Poissonian, albeit with a continuously changing average degree $\braket{k}$, plotted in the inset.

%

%
\subsection{Effect of dissipation on the universality class}
%
Dissipation is an essential ingredient of the SOC mechanism. It typically introduces characteristic length and time scales and thus strict scale invariance is lost. In the context of branching processes, for example, non-conservation of the particle number leads to a self-organization into an attractor state that is not critical but \textit{subcritical}, leading to an exponentially truncated distribution of avalanches \cite{lauritsen1996self}. As a result the distribution functions of avalanches no longer decay as a pure power law, but rather as
\begin{align}
    P(t) &\sim t^{-\tau_t} \, h_t(t/t_c),\nonumber\\
    P(s) &\sim s^{-\tau_s} \, h_s(s/s_c),\label{eq:cut-off}\\
    P(a) &\sim a^{-\tau_a} \, h_a(a/a_c).\nonumber
\end{align}
where the $h_\mu(x)$ are cut-off functions with cut-off scales $x_c$ that grow with decreasing dissipation strength. Some authors have referred to this as \textit{quasicriticality} \cite{bonachela2009self}. It is also argued \cite{bonachela2009self} that a "loading mechanism" can counteract the dissipation by replenishing the lost particles after each avalanche, as is done in e.g. the OFC model of SOC \cite{lise1996transitions}.

As long as there are no observable differences for experimentally-relevant system sizes we will here not distinguish between \emph{quasi-critical} and  \emph{critical}, i.e. truly scale-free behavior. Instead we focus on the question whether or not dissipation modifies the universality class in the spreading process of Rydberg facilitation, i.e. if the critical exponents $\tau_\mu$ in eq.\eqref{eq:cut-off} are modified. In particular we will explore by numerial simulations if dissipation acts as a relevant perturbation in the renormalization sense. Since our atom loss mechanism is conceptually very similar to that of the generalized epidemic process (GEP) or more generally of dynamical percolation (DyP), the presence of dissipation may change the critical behavior to that of the DyP universality class and  
we will compare our results with the corresponding predictions. In the GEP and DyP models, an individual's probability to be infected for the first time and that of all subsequent infections differ, the latter being set to zero in the extreme (GEP) case, which is referred to as perfect immunization. For the more general case of reduced repeated infection probability the phase transition is part of the DyP universality class \cite{hinrichsen2008book, grassberger1997spreading}. The three-dimensional case of DyP is characterized by the critical exponents \cite{hinrichsen2008book}
\begin{align}\label{eqn:dyp_aspt_exp_values}
\beta = 0.417,
&&
\nu_\parallel = 1.169,
&&
\nu_\perp = 0.875,
\end{align}
which via \eqref{eqn:tau_of_beta} then result in
\begin{align}\label{eqn:dyp_aval_exp_values}
\tau_a \approx 1.159,
&&
\tau_s \approx 1.123,
&&
\tau_t \approx 1.357.
\end{align}

In our model of Rydberg facilitation, dissipation is controlled by the $b$ parameter. The value $b=0$ corresponds to a dynamics where all Rydberg atoms return to the ground state after an exponentially distributed time. The value $b=1$, however, leads to the guaranteed irreversible loss of this atom from a Rydberg state.

\begin{figure}[H]
    \centering
    \includegraphics[width=\columnwidth]{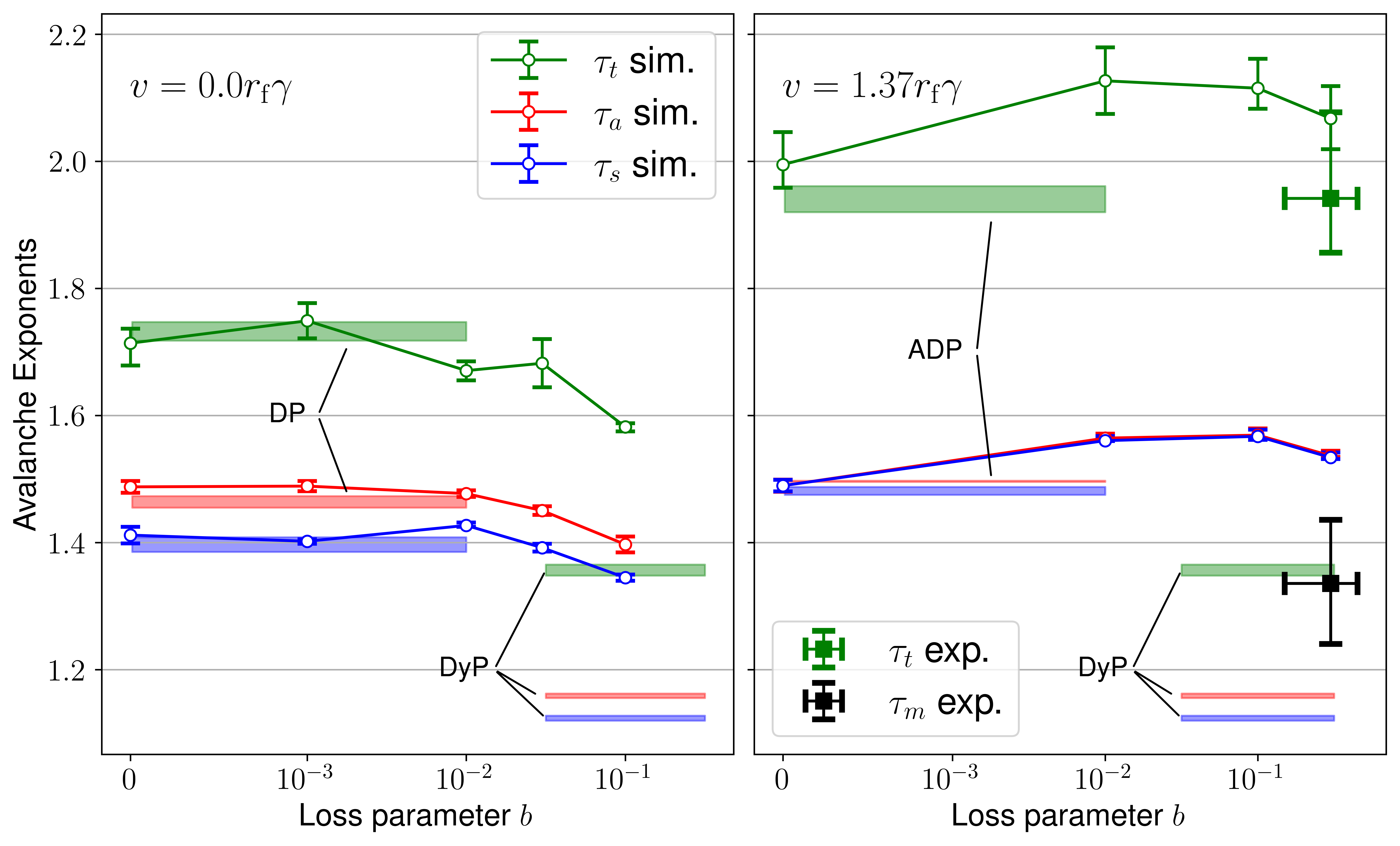}
    \caption{Finite-size extrapolated avalanche exponents over loss parameter $b$, for the cases of $v=0$ in the DP regime (left) and $v=1.37\rfac\decay$ for the ADP case (right). The red, blue and green bars show the expected exponents of area, size and time, respectively, for the universality class as stated. Note that in the case of the frozen gas for $b>0.1$ the distributions of area, size and time become increasingly characterized by an early exponential cutoff, resulting in a poor power-law fit, which is why we focus on $b<0.1$.}
    \label{fig:exponents_over_decay}
\end{figure}
%

To address the effect of dissipation we simulate the avalanches in the system with varying values of $b>0$. Generically we find a power law scaling over 1-2 orders of magnitude truncated
by an exponential cut-off function, eq.\eqref{eq:cut-off}. We extract the power-law exponents and perform finite-size extrapolations. The result can be seen in Fig.~\ref{fig:exponents_over_decay}. We observe that at $v=0$ for loss parameters $b<0.01$ we do not find a difference in the values of the exponents larger than our uncertainty. For larger $b$, the found exponents diminish in magnitude, especially the time exponent. For the case of $v=1.37\rfac\decay$, which is approximately the gas velocity in the experiment, we find that the avalanche exponents follow a non-monotonous behavior of increase for smaller $b$ and decreasing in magnitude again for larger $b$. Importantly, this implies that the avalanche exponents found in the experiment, which we also show in Fig.~\ref{fig:exponents_over_decay}, do not incur a significant additional systematic error based on non-vanishing $b-$values. In Fig.~\ref{fig:exponents_over_decay} we also show the predicted values for DyP. We observe that neither the simulated nor the experimentally measured values agree with those of DyP, showing that despite the conceptual similarity Rydberg facilitation cannot be simply pictured as an epidemic spreading with immunization. We have checked that this deviation does not result from the network structure set by the random atom positions in a gas by repeating our avalanche simulations on a regular 2D lattice of atoms with nearest-neighbor facilitation. Here, we also find no agreement with DyP exponents for $b\to 1$. We speculate that the difference might be found in the infection mechanism: In lattice models of DyP, an infected site passes on the infection to adjacent sites with a given probability, but decays after a single step of discrete time \cite{jimenez2003epidemic, grassberger1997spreading}, whereas in our model, both infection and decay occur probabilistically according to certain rates.

\section{Conclusion}

We studied the critical properties of excitation growth in a gas of atoms under conditions of Rydberg facilitation, which
represents an experimentally accessible model system for a spreading process on a random and dynamical network. 
In particular we determined the power-law exponents of the distribution of avalanches at the critical point of the absorbing-state phase transition (ASPT) both from numerical simulations and
experimental measurements. These exponents
can be related to the full set of critical exponents of the non-equilibrium phase transition and thus uniquely determine the universality class eq.\eqref{eqn:tau_of_beta}. In a previous  theoretical work we have provided numerical evidence that 
with increasing rms velocity of the atoms in the gas, the
character of the ASPT smoothly changes from directed percolation (DP) universality through different classes of anomalous directed percolation (ADP) to eventually mean-field (MF) behavior, which also explained previous experimental observations \cite{helmrich2020signatures}.
The velocity-dependent crossover was interpreted using a phenomenological model that mapped the Rydberg facilitation in the gas of moving atoms, resembling a dynamic network to an excitation spreading process on a random static network with Lévy-flight tails in the distribution of excitation distances \cite{brady2024anomalous}. Our simulations together with experimental results confirm that the avalanche distribution exponents follow the predictions obtained from the phenomenological model in \cite{brady2024anomalous} using the mapping relations \eqref{eqn:tau_of_beta} combined with previous results on the velocity dependence of the Lévy-flight parameter at the ASPT. Furthermore our work has given, to the best of our knowledge, the first experimental evidence of ADP universality.

We also investigated the network structure and its effects on the SOC mechanism. Since SOC on dynamical networks has been little researched and is as of yet poorly understood, we first characterize the dynamical properties of the underlying network structure of atoms in mutual facilitation distance as a function of gas velocity and quantify the number of dynamical connections. Secondly we investigated the influence of dissipation, important for the SOC, on the critical behavior. For the frozen gas, we first verified that the Erdős–Rényi character of the network is unaffected by decay. We then analyzed the influence of decay on the critical scaling in the DP and ADP regimes. While we cannot make any claims about the presence or absence of true critical behavior over arbitrary time and length scales, the observed power laws in the avalanche distributions over extended parameter ranges even in the presence of losses are consistent with at least quasi-critical behavior. The question we addressed instead was if losses modify the universality class of the ASPT in Rydberg facilitation, which could be the case if dissipation was a relevant perturbation in the renormalization sense. For a frozen gas we find that below a minimal dissipation probability we cannot detect a measurable influence on the scaling exponents. At stronger dissipation we see that the avalanche exponents are slightly reduced in magnitude. However, despite the conceptual similarities, we do not obtain exponents belonging to the dynamical percolation (DyP) universality class.

\subsection*{Acknowledgments}
The authors thank Fabian Isler for fruitful discussions.
Financial support from the DFG through SFB TR 185, Project No. 277625399, is gratefully acknowledged. The authors also thank the Allianz f\"ur Hochleistungsrechnen (AHRP) for giving us access to the “Elwetritsch” HPC Cluster. This work was also supported by the Max Planck Graduate Center with the Johannes Gutenberg-Universit\"at Mainz (MPGC) and the Quantum Initiative Rhineland-Palatinate (QUIP) and the Research Initiative Quantum Computing for Artificial Intelligence QC-AI. W.R. acknowledges support by the German Ministry of Education and Research (BMBF) for BIFOLD (01IS18037A).

\subsection*{Author contributions}

S.O. performed the numerical calculations and the analysis of critical exponents with support from D.B. The experiment was performed by J.B., P.M. and D.B. guided by T.N. and H.O.
W.R. and J.S.O. helped in numerical implementation and optimization. J.S.O. additionally helped in supervising the theoretical work.
M.F. and T.N. conceived the project and supervised the theoretical (M.F.) and experimental parts (T.N.) of the project, respectively. S.O., D.B. and M.F. wrote the initial version of the paper. All authors discussed the results and contributed to the writing of the final manuscript.

\newpage
\appendix

\section{Rate Equation Modeling}
\label{sect:appendix_rate_equations}

All numerical data is obtained using fixed-time step Monte-Carlo simulations \cite{barlett2009differences} of classical rate equations in the high dephasing limit. It has been shown that in this limit dynamics become effectively classical and can therefore be described by classical Monte-Carlo simulations to a high degree of accuracy \cite{levi2016quantum}. 

For atom~$i$, the excitation probability is given by the projection operator onto the Rydberg state $\ket{R}_i$, i.e. ${\srr_i = \ket{R}_i \bra{R}_i}$. Using the Lindblad master equation, given by eq.~\eqref{eq:master_equation}, we can formulate a set of differential equations for the ground state $\ket{G}_i$, Rydberg state $\ket{R}_i$, and inert state $\ket{0}_i$ of the $i$-th atom. After adiabatic elimination of coherences, e.g. ${\frac{d}{dt} \sigma^{gr}_i = 0}$, (where $\sigma^{gr} = \vert G\rangle\langle R\vert$) we receive the rate equations \cite{brady2024griffiths, brady2023mean}
\begin{align}
    \frac{d}{dt} p_r^{(i)} &=  
    -(\gamma_\mathrm{stim} + \gamma_\mathrm{spont}) p_r^{(i)}
    + \gamma_\mathrm{stim} p_g^{(i)},
    \\
    \frac{d}{dt} p_g^{(i)} &= -\gamma_\mathrm{stim} p_g^{(i)} + (\gamma_\mathrm{stim} + (1 - b) \gamma_\mathrm{spont}) p_r^{(i)},
    \\
    \frac{d}{dt} p_0^{(i)} &= b \gamma_\mathrm{spont} p_r^{(i)}.
\end{align}
Here $\gamma_\mathrm{spont}$ corresponds to the spontaneous decay rate and $\gamma_\mathrm{stim}$ corresponds to the stimulated (de-)excitation rate. Explicitly, the stimulated rate reads
\begin{align}
    \gamma_\mathrm{stim} = \frac{2 \Omega^2 \gamma_\perp}{\gamma_\perp^2 + \Delta^2 
    \big(
        \sum_{\substack{j \neq i \\ j \in \Sigma}} \frac{\rfac^6}{r_{ij}^6} - 1
    \big)^2
    },
\end{align}
where $\Sigma$ corresponds to the subset of atoms in the Rydberg state. 

We initiate $N$ atoms with random positions in a cubic simulation box with length $L$ and periodic boundary conditions. Velocities are sampled from the Maxwell-Boltzmann distribution, i.e. a Gaussian in each direction, with the most probable velocity $v$. Furthermore, we dynamically adjust the (fixed) time step length depending on the facilitation rate with ${\gfac \, dt = \frac{1}{10}}$. In order to receive good avalanche fits, we use approximately 200,000 trajectories per parameter set.

\section{Finite-Size Expansion of Avalanche Data}\label{appendix_fse}

\begin{figure}[t]
    \centering
    \includegraphics[width=1.\columnwidth]{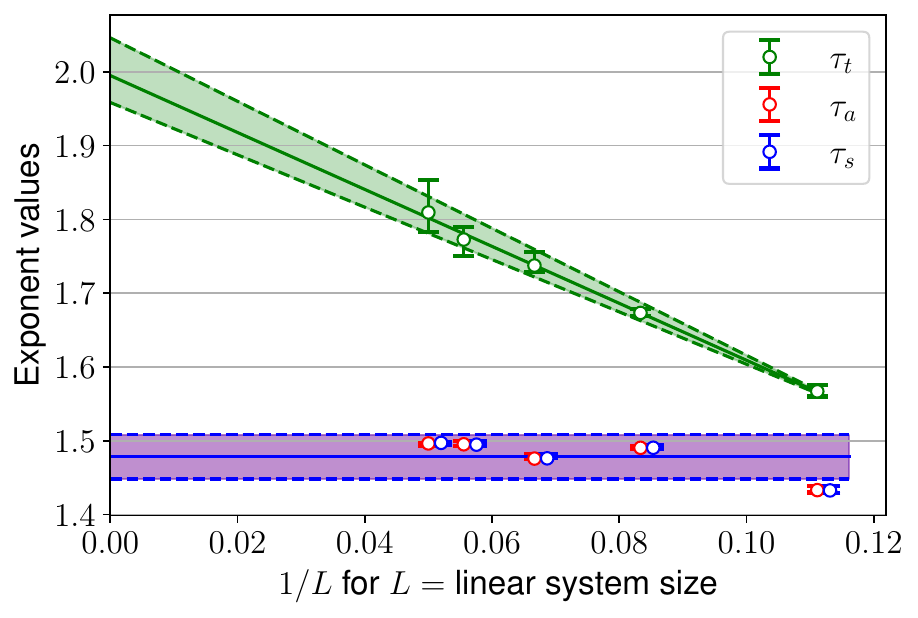}
    \caption{Finite-size expansion of the avalanche exponents. The shaded area corresponds to the uncertainty interval. Size values are shifted slightly along the x-axis for better visibility.}
    \label{fig:exponents_fse_showcase}
\end{figure}
%
All of the simulation exponent values reported in this paper were obtained using a system size extrapolation. We fit a linear function $f(1/L)$, where $L$ is the linear system size, to the exponent values for different system sizes and then extrapolate to $f(0)$. In Fig.~\ref{fig:exponents_fse_showcase} we show the extrapolation for the case of $v=1.37\rfac\decay$ in the ADP phase. For the time exponent we see a clear finite-size scaling with an increasing value for larger systems, whereas the area and size exponents do not show a clear trend over system size as well as a much smaller variation. The shaded areas correspond to the uncertainty region. 

\section{Determining the critical point}\label{appendix_critical_point_estimation}
\subsection{Numerical Simulations}
\begin{figure}[t]
    \centering
    \includegraphics[width=1.\columnwidth]{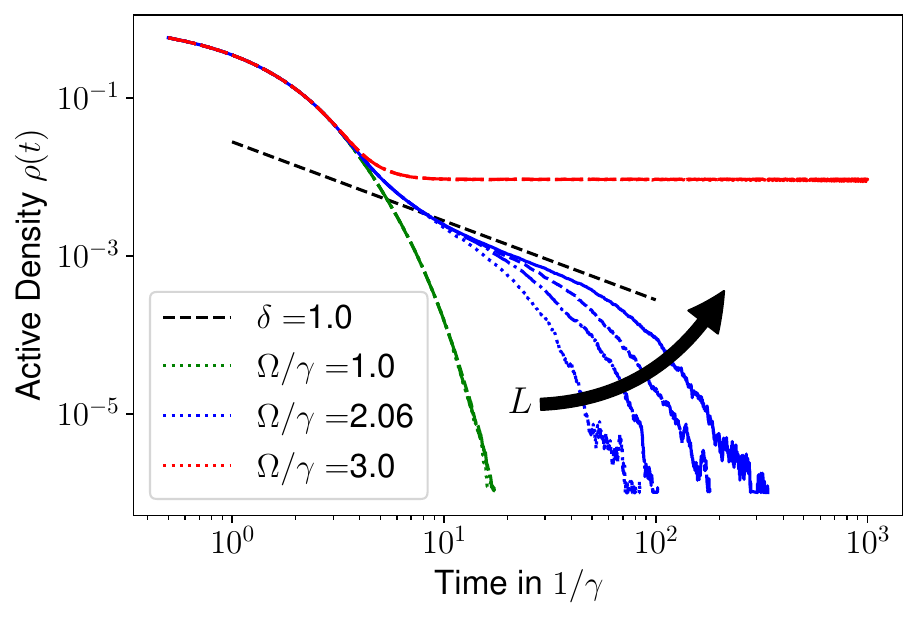}
    \caption{Finite-size expansion of the active density decay method of finding $\Omega_c$ for $v=10\rfac\decay$. For all values of $\Omega$, we show the averaged decay data for $L\in\{5, 7, 9\}\rfac$, for $\Omega/\decay=2.06$ we also show $L=12\rfac$. The linestyles with increasing system size are: dotted, dash-dotted, dashed, solid.}
    \label{fig:add_fse_showcase}
\end{figure}

Finding the correct critical point $\Omega_c$ is essential in obtaining power-law distributed avalanches. We determine $\Omega_c$ in our numerical simulations by starting from the fully inverted state (all atoms in the Rydberg state) and considering the decay process as a function of the system size. In the absorbing phase, the decay is exponential and shows no strong dependence on system size. In the active phase, the Rydberg density approaches a constant. For values of $\Omega$ close to the critical point a regime with power-law decay emerges for intermediate time scales, where the precise value of $\Omega_c$ is then obtained by fitting a power-law function with the exponent $\delta=\frac{\beta}{\nu_\parallel}$ to increasing system sizes $L$ as shown in Fig.~\ref{fig:add_fse_showcase}. Note that for the absorbing as well as the active case the curves for all system sizes lie on top of each other. Also see \cite{munoz2010griffiths} for more details. We find $\Omega_c/\decay\approx 3.40$ for the frozen gas, $\Omega_c/\decay\approx 2.325$ for the ADP regime ($v=1.37\rfac\decay$) and $\Omega_c/\decay\approx 2.06$ for the mean-field regime ($v=10\rfac\decay$).

\subsection{Experimental Data}

\begin{figure*}[t]
    \includegraphics[width=\textwidth]{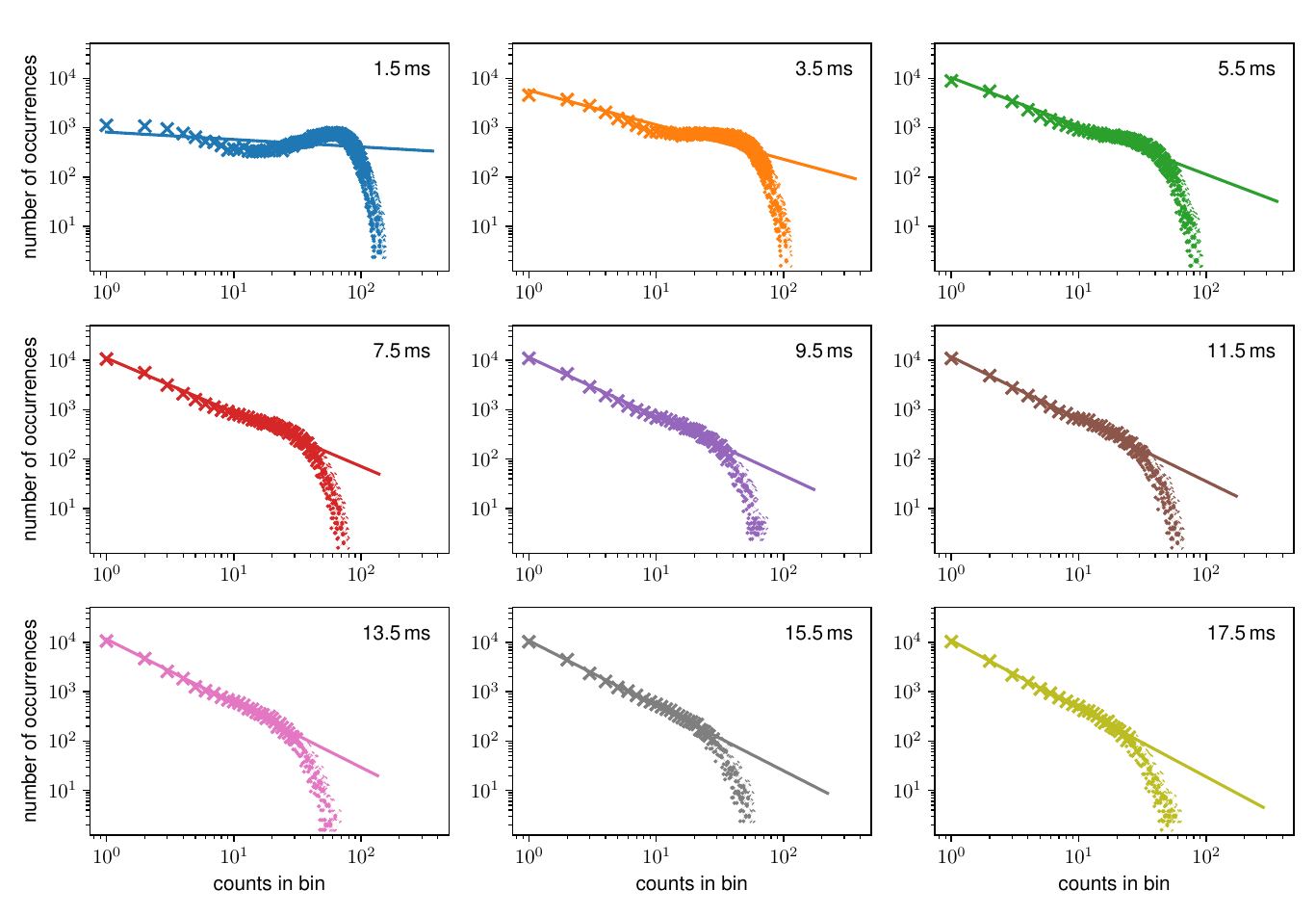}
    \caption{The activity distribution of the experimental system at different time intervals. In each subplot, the activitiy distribution for a different time interval of $\Delta t=\SI{3}{\milli\second}$ is shown, centered as indicated in the corner of the plot. The distribution of count numbers in \SI{50}{\micro\second} bins is shown. The fitted lines serve as a guide to the eye to evaluate, whether the distribution is active. For early times up until the interval of $\SI{6}{\milli\second} - \SI{9}{\milli\second}$ we find a "bump" at large activity sizes, which is indicative of the active phase. We estimate that in the subsequent time windows of $\SI{8}{\milli\second} - \SI{13}{\milli\second}$ the critical point is reached.}
    \label{fig:experiment_critical_point_estimate}
\end{figure*}

The extracted exponents of the power-law distributed avalanches strongly depend on the time windows in which they are evaluated. It is a challenging task to determine the correct point in time at which the critical point is reached. 
In the actives phase, the system typically forms a single large cluster of Rydberg excitations where the total number of excited atoms is effectively limited by the size of the system.
The impact of this size limitation can be observed as complex non-algebraic behavior in the activity distribution.
In Fig. \ref{fig:experiment_critical_point_estimate} we show histograms of the count numbers in a single evaluation bin for different time windows. As the number of ions in a fixed time $t_\mathrm{bin}$ is proportional to the number of Rydberg atoms times the decay rate, this can be understood as a measure for the activity of the system.
In the active phase, which is our starting point, the histograms show a characteristic activity bump for large counts. We estimate that the active phase ends when the distribution shows no residuals of such an activity bump clearly visible in the first evaluation frame. In our experiment, we estimate that this is the case in between \SI{8}{\milli\second} and \SI{13}{\milli\second}, i.e. time window centers of \SI{9.5}{\milli\second} and \SI{11.5}{\milli\second}.

We note that other, more indirect, ways to estimate the critical point are possible, e.g. by exploiting scaling relations of avalanche shapes \cite{Sethna2001crackling}. Moreover, the critical point may be reached after slightly different times in each individual experimental realization, given that the loss process is stochastic. 
\newpage

\bibliography{references}

\end{document}